\documentclass{llncs}
\usepackage{amsmath, amssymb}

\begin{document}
\pagestyle{plain}
\title{New Class of Optimal Frequency-Hopping Sequences by Polynomial Residue Class Rings }

\author{Wenping Ma \inst{1} and Shaohui Sun\inst{2}}
\institute{$^{1}$ National Key Lab. of ISN, Xidian University , Xi'an 710071, P.R.China\\
\email{wp\_ma@mail.xidian.edu.cn} \\
$^{2}$ Datang Mobile Communications Equipment Co.,Lid,Beijing
100083, P.R.China }

\maketitle

\begin{abstract}
\noindent In this paper, using the theory of polynomial residue
class rings, a new construction is proposed for frequency hopping
patterns having optimal Hamming autocorrelation with respect to the
well-known $Lempel$-$Greenberger$ bound.  Based on the proposed
construction, many new $Peng$-$Fan$ optimal families of frequency
hopping sequences are obtained. The parameters of these sets of
frequency hopping sequences are new and flexible.
\bigskip

\end{abstract}

\section{Introduction}

Let $\mathcal{F}=\{f_0,f_1,\cdots,f_{l-1}\}$   be a set of available
frequencies, called an $alphabet$. Let $\textit{S}$ be the set of
all sequences of length  $\nu$ over $\mathcal{F}$. Any element of
$\textit{S}$  is called a frequency-hopping sequence of length $\nu$
over $\mathcal{F}$. Given any two frequency hopping sequences
$X,Y\in \textit{S}$, we define their Hamming correlation $H_{X,Y}$
to be
$$H_{X,Y}(t)=\sum^{\nu-1}_{i=0}h[x_i,y_{i+t}],0\leq t<\nu,$$
where $h[a,b]=1$ if $a=b$, and 0, otherwise, and all operations
among the position indices are performed modulo $\nu$. For any
distinct
$X,Y \in \textit{S}$, we define\\
$$H(X)=\max_{1\leq t< \nu}{\{H_{X,X}(t)\}}$$
$$H(X,Y)=\max_{0\leq t< \nu}{\{H_{X,Y}(t)\}}$$
$$ M(X,Y)= \max{\{H(X),H(Y),H(X,Y)\}}.$$
Lempel and Greenberger\cite{2} developed the following lower bound
for $H(X)$. \textbf{Lemma 1:} For every frequency hopping sequence X
of length $\nu$ over
an alphabet of size $l$ , we have\\
$$ H(X)\geq{\biggl\lfloor
{(\nu-\varepsilon)(\nu+\varepsilon-l)\over{l(\nu-1)}}\biggr\rfloor}$$\\
where $\varepsilon$  is the least nonnegative residue of $\nu$
modulo $l$.\\
 \textbf{Corollary 1(6):} For any single frequency hopping sequence
of length $\nu$ over an alphabet of size $l$ , we have
\begin{displaymath}
H(X)\geq\left \{\begin{array}{ll}$$k,if\thinspace\thinspace \nu \neq
l$$\\$$0,if\thinspace\thinspace \nu= l$$\end{array}\right.
\end{displaymath}
where $\nu = kl+\varepsilon$, $0\leq \varepsilon<l$.

Let $\Gamma$ be a subset of $\textit{S}$ containing $N$ sequences.
We define the maximum nontrivial Hamming correlation of the sequence
set $\Gamma$ as
$$M(\Gamma)=\max\{\max_{X\in\Gamma}H(X),\max_{X,Y\in\Gamma,X\neq Y}H(X,Y)\}$$
$$H_{a}(\Gamma)=\max_{X\in\Gamma}H(X)$$
$$H_{c}(\Gamma)=\max_{X,Y\in\Gamma,X\neq Y}H(X,Y)$$
\verb+   +Throughout this paper, we use $(\nu,N,l,\lambda)$ to
denote a set of $N$ frequency hopping sequences $\Gamma$ of length
$\nu$ over an alphabet of size $l$, where $\lambda=M(\Gamma)$ .

Peng and Fan\cite{3} developed the following bound on
$H_{a}(\Gamma)$and $H_{c}(\Gamma)$,
which take into consideration the number of sequences in the family.\\
\textbf{Lemma 2:}For any family of frequency hopping sequences
$\Gamma$ , with length $\nu$, an alphabet of size $l$ , and
$|\Gamma|=N$ , we have
$$(\nu-1)NH_a(\Gamma)+(N-1)N\nu H_c(\Gamma)\geq 2I\nu N-(I+1)Il$$
where $\verb+ + \displaystyle I=\lfloor{\nu N\over{l}}\rfloor.$\\
\textbf{Lemma 3(6):} For any pair of distinct frequency hopping
sequences $X,Y$, with $|\mathcal{F}|=l$, we have
$$M(X,Y)\geq{{4I\nu-(I+1)Il}\over{4\nu-2}}$$
where $2\nu=Il+r$ and $0\leq r<l$.

\begin{definition}
(1) A sequence   $X\in\textit{S}$ is called optimal if the
$Lempel$-$Greenberger$ bound in Lemma 1 is met.\\
 (2) A subset
$\Gamma\subset\textit{S}$  is an optimal set if the $Peng-Fan$
bound in Lemma 2 is met.\\
(3) Any pair of distinct frequency hopping sequence $\{X,Y\}\subset
\textit{S}$ constitute a $Lempel$-$Greenberger$ optimal pair of
frequency hopping sequences if the bound in Lemma 3 is met .
\end{definition}

Lempel and Greenberger\cite{2} defined optimality for both single
sequences and sets of sequences in other ways. A set of frequency
hopping sequences meeting the $Peng$-$Fan$ bound in Lemma 2 must be
optimal in the $Lempel$ and $Greenberger$ sense.

In modern radar and communication systems, frequency hopping
spread-spectrum techniques have been popular, such as frequency
hopping code division multiple access and ``Bluetooth"
technologies\cite{7,8}.

The objective of this paper is to present a new method to construct
new family of frequency hopping sequences. Both individual optimal
frequency-hopping sequences and optimal families of frequency
hopping sequences are presented.

\section{Polynomial Residue Class Rings Preliminary}
In the following, we introduce in brief polynomial residue class
rings preliminary. For details on polynomial residue class rings, we
refer to \cite{1}

\begin{definition} Let $p$ be a prime, $GF(p)$ be a finite field,
$GF(p)[\xi]$ be the ring of all polynomials over $GF(p)$, and
$\omega(\xi)$ be an irreducible polynomial of degree $m$ over
$GF(p)$, where $m \geq 1$. Then $\Re$ is defined as the quotient
ring generated by $\omega(\xi)^k$ in $GF(p)[\xi]$, $k\geq 1$ .
$$\Re = GF(p)[\xi]\Bigl{/}(\omega(\xi)^k)$$
\end{definition}

We have a natural homomorphic mapping, $\mu$ from $\Re$ to its
residue field $F = GF(p)[\xi]\Bigl{/}(\omega(\xi))$. Define
$\mu:\Re\rightarrow F$ by $\mu(a)= a\verb+ + mod\verb+ +
\omega(\xi)$ . It is easy to verify that the elements in the set
$\{1,\omega(\xi),\omega^2(\xi),\cdots,\omega^{k-1}(\xi)\}$ are
linearly independent over $F$ and hence constitute a basis of $\Re$
over $F$. Thus any element $a\in\Re$ can be represented uniquely as
$$a=a_0+a_1\omega(\xi)+\cdots+a_{k-1}\omega_{k-1}(\xi),a_i\in F,i=0,1,2,\cdots,k-1.$$
Thus $\Re$ can be written as
$$\Re=F+F\omega+F\omega^2+\cdots+F\omega^{k-1}\verb+     +(1)$$
The group of units $\Re^*$ of $\Re$ is given by the direct product
of two group $G_{PRC}$ and $G_{PRA}$, $\Re^*=G_{PRC}\times G_{PRA}$
, where $G_{PRC}$  is a cyclic group of order $p^m-1$ and $G_{PRA}$
is an Abelian group of order $p^{m(k-1)}$.\\
\textbf{Lemma 4:}The set $\{G_{PRC},0\}$ is isomorphic to residue
field $F$ and is also a subspace of $\Re$. Thus the set
$\{G_{PRC},0\}$ is a subring of $\Re$ .

From now on, we will omit the indeterminate $\xi$ from the
representation.

Let $\Re[x]$ be the ring of polynomials over $\Re$ . We extend the
homomorphic mapping $\mu$  on $\Re$  to polynomial reduction mapping
:

$\hat{\mu}:\Re[x]\rightarrow F[x]$ in the obvious way

\makeatletter
\newcommand{\Extend}[5]{\ext@arrow 0099{\arrowfill@#1#2#3}{#4}{#5}}
\makeatother
$$ f(x)=\sum^r_{i=0}a_ix^i\Extend{-}{-}{>}{}{\displaystyle\hat{\mu}\hspace{1em}}\sum^r_{i=0}\mu(a_i)x^i $$
A polynomial $f(x)\in \Re[x]$ is a basic irreducible if $\mu(f(x))$
is irreducible in $F[x]$; it is monic if its leading coefficient is
1.\\

\begin{definition} The Galois ring of $\Re$ denoted as
$GR(\Re,r)$ is defined as $\Re[x]\Bigl{/}(f(x))$, where $f(x)$ is a
basic monic irreducible polynomial of degree $r$ over $\Re$ .
\end{definition}

The group of units of $GR(\Re,r)$ denoted by $GR^*(\Re,r)$ is given
by a direct product of two groups:
$$GR^*(\Re,r)=G_C\times G_A$$
where $G_C$ is a cyclic group of order $p^{mr}-1$  and $G_A$ is an
Abelian group of order $p^{m(k-1)r}$ . On the lines of Lemma 4, it
is easy to show that the set $\{G_C,0\}$is a field of order $p^{mr}$
. This is denoted by $GF(p^{mr})$. Thus like the representation (1)
for $\Re$, we have
$$GR(\Re,r)=GF(p^{mr})+\omega GF(p^{mr})+\omega^2
GF(p^{mr})+\cdots+\omega^{k-1}GF(p^{mr}),$$ hence, any element
$\alpha\in GR(\Re,r)$ can be uniquely expressed
as \\
\\
$\alpha =
\alpha_0+\omega\alpha_1+\omega^2\alpha_2+\cdots+\omega^{k-1}\alpha_{k-1}$,$\alpha_i\in
GF(p^{mr})$, $i=0,1,\cdots,k-1.$\verb+      +(2)\\
\\
The elements of  $G_A$ are of the form $1+\omega(x)A'$ , where
$A'\in GR(\Re,r)$. From (2), the elements of $G_A$ are given by the
set
$$\{(1+\omega\gamma),\gamma=\gamma_0+\omega\gamma_1+\cdots+\omega^{k-2}\gamma_{k-2},\gamma_i\in GF(p^{mr}\}\verb+                  +(3)$$
The Galois automorphism group of $GR(\Re,r)$ over its intermediate
subring $GR(\Re,s)$, where $s$ divides $r$ is cyclic of order
$(r/s)$ generated by the Frobenius map $\sigma^s$ defined by
$$\sigma^s(\alpha)=(\alpha_0)^{p^s}+(\alpha_1)^{p^s}\omega+\cdots+(\alpha_{k-1})^{p^s}\omega^{k-1}$$
where $\alpha$ is as in (2). When $s=1$ , the above Frobenius map
generates Galois group over $\Re$. Using the automorphisms given
above, we define below generalized trace functions which map
elements of $GR(\Re,r)$ to its intermediate subrings $GR(\Re,s)$
where $s$ divides $r$. They are given by
$$Tr^r_s(\alpha)=\sum^{(r/s-1)}_{i=0}[(\alpha_0)^{p^{si}}+(\alpha_1)^{p^{si}}\omega+(\alpha_2)^{p^{si}}\omega^2+\cdots+(\alpha_{k-1})^{p^{si}}\omega^{k-1}]$$
where $\alpha\in GR(\Re,r)$.The above trace function is the
generalization of trace function defined for finite fields. Like
their counterparts in finite fields, the trace functions satisfy the
following properties:
$$Tr^r_s(\alpha)=Tr^r_s(\sigma^{si}(\alpha)),for\verb+ + all\verb+ +
i.$$ $Tr^r_s(a\alpha+b\beta)=aTr^r_s(\alpha)+bTr^r_s(\beta)$;
$\forall
a,b\in GR(\Re,s)$ and $\forall \alpha,\beta\in GR(\Re,r)$.\\
\\
For any fixed $b$ of $GR(\Re,s)$, the equation $Tr^r_s(\alpha)=b$,
has exactly $p^{mk(r-s)}$ solutions in $GR(\Re,r)$.
$$Tr^r_1(\alpha)=Tr^s_1(Tr^r_s(\alpha)).$$
\begin{theorem}\cite{1}
Every  $m$-sequence over $\Re$ has a unique trace representation
given by $\{s^{\gamma}_i\}^{\infty}_{i=0}=Tr^r_1(\gamma\alpha^i)$,
where $\gamma \in GR(\Re,r)$ and $\alpha$ is a primitive root of
$f(x)$ and belongs to $G_C$.

We shall denote $S^*(f)$ as the set of sequences which contains not
all zero divisors. By using the structure of group of units
$GR^*(\Re,r)=G_C \times G_A$ and (3), all $m$-sequence in $S^*(f)$
are given by the set
$$\{(s^{\gamma}_i)^{\infty}_0,\gamma=(1+\omega(\gamma_0+\gamma_1\omega+\cdots+\gamma_{k-2}\omega^{k-2})), where\verb++
\gamma_j\in GF(p^{mr}),j=0,1,\cdots,k-2\}.$$
\end{theorem}
\begin{definition}Let $\alpha\in GR(\Re,r)$ as in (2) be
equal to
$\alpha_0+\alpha_1\omega+\cdots+\alpha_{k-1}\omega^{k-1}$,$\alpha_i\in
GF(p^{mr})$. Then, let $M_{\alpha}$ be a matrix over $F$ of
dimension $r\times k$ formed by placing together $k$ elements
$\alpha_0,\alpha_1,\cdots,\alpha_{k-1}$ as columns of M. Then the
rank number $\kappa(\alpha)$ of $\alpha$ is defined as the rank of
matrix $M_{\alpha}$ over $F$.
\end{definition}
\begin{definition} Given a sequence $\mathcal{S}$ and an
element $s$ of $\Re$, we define $W_s(\mathcal{S})$ as the number of
occurrences of the element s in $\mathcal{S}$ within its one period
length.
\end{definition}

\begin{theorem} \cite{1}
Let $\{s^{\gamma}_i\}^{\infty}_0$ be an $m$-sequence with
$\kappa(\gamma)=\rho$. Then,$W_{0^{k}}(s^{\gamma})=p^{m(r-\rho)}-1$,
and $W_s(s^{\gamma})=p^{m(r-\rho)}$, for $s\neq 0^{k}$.
\end{theorem}
\begin{definition}The Trace Image of an  $m$-sequence,
$s^{\gamma}$ is defined as the set of distinct elements in
$s^{\gamma}$. The cardinality of the Trace Image is given by
$p^{m\rho}$.
\end{definition}

\section{New Optimal Frequency Hopping Sequences from Residue Class Rings}
Let$\verb+ + q=p^m,$ $z$\verb+ + is a positive integer satisfying
$\verb+ +z|(q-1), n= {{q^r-1}\over{z}}$, $r$ is a positive integer,
in this paper, we suppose $ gcd(\frac{q^{r}-1}{q-1},z)=1$,  $\alpha$
be a primitive generator of $G_C$ present in $GR^*(\Re,r)$,$\gamma
\in G_{A}$ with $\kappa(\gamma)=\rho$.

Let $s$ be an integer with $gcd(s,q^r-1)=1$ , and define
$\beta={\alpha}^{z s}$ . It is easy to check that the minimal
positive integer $d$ satisfying ${\beta}^{q^d-1}=1$ is $r$ , thus
$1,\beta,\beta^2,\cdots,\beta^{r-1}$ is linear independent over
$G_{PRC}$.

We define the following sequence:
$$s^{(\gamma,g)}_i=Tr^r_1(\gamma g\beta^i),i=0,1,\cdots,k,\cdots,g\in
G_C$$ It is easy to check that
$s^{(\gamma,g)}_i=s^{(\gamma,g)}_{i+n}$ , then
$(s^{(\gamma,g)}_i)^{\infty}_{0}$ is a sequence of period $n$.

We define the following sequences set: \\
$$\Gamma=\{(s^{(\gamma,\alpha^{sk})}_i)^{\infty}_{0}:0\leq k<z\}
\verb+                +(4)$$\\
It is obvious that $|\Gamma|=z$.

\begin{definition}Two
sequences ${(s^{(\gamma,g)}_i})_{0}^{\infty}$ and
$(s^{(\gamma,g')}_i)^{\infty}_{0}$ are called projectively
cyclically equivalent if there exist an integer $t$  and a nonzero
scalar
$\lambda\in G_{PRC}$ \\
$$s^{(\gamma,g)}_i=\lambda s^{(\gamma,g')}_{i+t},  i=0,1,2,\cdots. \verb+                   +(5)$$ \\
\end{definition}
We wish to count the number of inequivalent in $\Gamma$  using (5)
as the definition of equivalence.
\begin{theorem}
For any two sequences  $(s^{(\gamma,g)}_i)^{\infty}_{0}$ and
$(s^{(\gamma,g')}_i)^{\infty}_{0}$  belonging to $\Gamma$, they are
projectively cyclically equivalent.
\end{theorem}
\textbf{Proof:}Formula (5) can be written as
$$Tr^r_1(\gamma g\beta^i)=Tr_1^r(\gamma\lambda g'\beta^{(t+i)}),i\geq 0$$
$$Tr^r_1[\gamma(g-\lambda g'\beta^t)\beta^i]=0,i\geq 0$$
It follows that Formula (5) is equivalent to
$${g\over{g'}}=\lambda\beta^t\verb+                               +(6)$$
The set of elements in $G_C$ of the form $\lambda\beta^t$  where
$\lambda\in G_{PRC}$ is a subgroup of the multiplicative group of
nonzero elements of $G_C$. What (6) says is that $g$ and $g'$  are
equivalent if and only if  $g$ and $g'$  lie in the same coset of
this subgroup. It follows that the number of inequivalent $g$'s is
equal to the number of such cosets, viz.
$$N_{1}={(q^r-1)\over{|G|}},$$
where $G$  is the subgroup of elements of the form
$\{\lambda\beta^i\}$. It remains to calculate $|G|$. Now $G$ is the
direct product of the two groups $G_{PRC}$ and
$A=\{1,\beta,\cdots,\beta^{n-1}\}$. From elementary group theory we
have
$$|G|={|A|\cdot|G_{PRC}|\over{|G_{PRC}\cap A|}}.$$
To calculate $|G_{PRC}\cap A|$ we note that this number is just the
number of distinct powers of  $\beta$, which are elements of
$G_{PRC}$ . But $\beta^i\in G_{PRC}$ if and only if
$\beta^{i(q-1)}=1$. Since $ord(\beta)=n$, this is equivalent to
$n|i(q-1),i.e,$
$${n\over{gcd(n,q-1)}}|i$$
Thus if we define
$$e=gcd(n,q-1)$$
$$d={n\over{e}}.$$

Because
$e=gcd(n,q-1)=gcd(\frac{q^{r}-1}{q-1}\frac{q-1}{z},z\frac{q-1}{z})$
and $\displaystyle gcd(z,\frac{q^{r}-1}{q-1})=1$, then
$\displaystyle e=\frac{q-1}{z}$.

We see that $\beta^i\in G_{PRC}$ iff $i=0,d,2d,\cdots,(e-1)d$, hence
$|G_{PRC}\bigcap A|=e$, and we have
$$|G|=n(q-1)\Bigl{/}e=q^{r}-1,$$
$$N_1=1.$$

\begin{theorem}
$$W_{0^{k}}((s^{(\gamma,g)}_{i})^{\infty}_{0})={q^{r-\rho}-1\over{z}}.$$
\end{theorem}
\textbf{Proof:}Let $1,\alpha^s,\cdots,\alpha^{s(z-1)}$  be a
complete set of representatives for the cosets of
$\{1,\beta,\cdots,\beta^{n-1}\}$ in the multiplicative group $G_C$.
Every nonzero element $\theta\in G_C$ can be written as
$\theta=\alpha^{si}\beta^j$ for a unique pair $(i,j),0\leq i\leq
z-1$,$0\leq j\leq n-1$. Now consider the following $z\times n$
array, which we call Array 1:\\
\\
$$\begin{array}{ccccc}
1&\beta & \beta^2 & \cdots &\beta^{n-1}\\
\alpha^s & \alpha^s\beta &\alpha^s\beta^2 & \cdots & \alpha^s\beta^{n-1}\\
\alpha^{2s} & \alpha^{2s}\beta & \alpha^{2s}\beta^2 & \cdots &
\alpha^{2s}\beta^{n-1}\\
\vdots & \vdots & \vdots & \vdots & \vdots \\
\alpha^{(z-1)s} & \alpha^{(z-1)s}\beta & \alpha^{(z-1)s}\beta^2 &
\cdots & \alpha^{(z-1)s}\beta^{n-1}
\end{array}$$
Now let $s_{ij}=Tr^r_1(\alpha^{is}\beta^j)$ and consider this
array,which we call Array 2:
 $$\begin{array}{ccccc}
 s_{00} & s_{01}& s_{02}& \cdots & s_{0(n-1)}\\
 s_{10} & s_{11}& s_{12}& \cdots & s_{1(n-1)}\\
 s_{20} & s_{21}& s_{22}& \cdots & s_{2(n-1)}\\
 \vdots & \vdots & \vdots & \vdots & \vdots \\
 s_{(z-1)0} & s_{(z-1)1}& s_{(z-1)2}& \cdots & s_{(z-1)(n-1)}\\
 \end{array}$$
 Since Array 2 is the ``trace" of Array 1, and since every nonzero element of $G_C$ appears exactly once in Array 1,
 It follows that $0$  appears exactly $q^{(r-\rho)}-1$  times in Array 2. Finally, since $N_1=1$ ,
 we know that every row of Array 2 can be obtained from the first row by shifting and multiplying by scalars.
 Thus 0 appears the same number of times in each row of Array 2. Since there are $z$ rows in the array,
 and 0 appears $q^{(r-\rho)}-1$  time altogether, each row contains exactly
 $\displaystyle{q^{r-\rho}-1\over{z}}$ 0.
 \begin{theorem}
 $\{s^{(\gamma,g)}_{i}\}^{\infty}_0$ is an optimal frequency hopping sequence with
 parameters $\displaystyle ({q^r-1\over{z}},q^{\rho},{q^{r-\rho}-1\over{z}}).$
 \end{theorem}
\textbf{Proof:}Because $\displaystyle {q^r-1\over{z}}=q^{\rho}\cdot
{q^{r-\rho}-1\over{z}}+{q^{\rho}-1\over{z}}$ , the conclusion
follows from Lemma 1 and Corollary 1.
\begin{theorem}
if $g,g'$ belong to distinct cyclotomic classes of order $z$  in
$G_C$ , then $((s^{(\gamma,g)}_i)^{\infty}_{0})$ and
${(s^{(\gamma,g')}_i)}^{\infty}_{0}$ constitute a
$Lempel-Greenberger$ optimal pair of frequency hopping sequences.
\end{theorem}
\textbf{Proof:} By Theorem 5,
$H_a((s^{(\gamma,g)}_i)^{\infty}_{0})=H_a((s^{(\gamma,g')}_i)^{\infty}_{0})$
. Now we compute the cross-correlation values of
$(s^{(\gamma,g)}_i)^{\infty}_{0}$ and
$(s^{(\gamma,g')}_i)^{\infty}_{0}$. From the definition of
$s^{(\gamma,g')}_i$, we know that for any $t\in\{0,1,\cdots,n-1\}$,
if we cyclically shift $s^{(\gamma,g')}_i$ to the left for $t$ time,
we obtain $s^{(\gamma,g')}_{i+t}=Tr_1^r(\gamma
g'\beta^t\beta^i),i=0,1,2,\cdots,$ then, by noting that
$s^{(\gamma,g)}_i-s^{(\gamma,g')}_{i+t}=Tr^r_1[\nu(g-g'\beta^t)\beta^i],i=1,2,\cdots$
. Since $g,g'$ are in distinct cyclotomic classes of order $z$ in
$G_C$ , $g-g'\beta^t$ can never be zero. It then follows from
Theorem 4 that
$$H_{(s^{(\gamma,g)}_i)^{\infty}_{0}, (s^{(\gamma,g')}_{i})^{\infty}_{0}}(t)=\displaystyle
{{q^{r-\rho}-1}\over{z}}.$$ For any $t\in\{0,1,\cdots,n-1\}$ .
Therefore we can conclude that
$H((s^{(\gamma,g)}_i)^{\infty}_{0},(s^{(\gamma,g')}_i)^{\infty}_{0})=\displaystyle
{q^{r-\rho}-1\over{z}}$ . We claim that
$(s^{(\gamma,g)}_i)^{\infty}_{0}$ and
$(s^{(\gamma,g')}_i)^{\infty}_{0}$ constitute a $Lempel-Greenberger$
optimal pair of frequency hopping sequences, if $g,g'$ belong to
distinct cyclotomic classes of order $z\geq 2$ in $G_C$. In fact,
for any two $q^{\rho}$-ary sequences
$(s^{(\gamma,g)}_i)^{\infty}_{0}$ and
$(s^{(\gamma,g')}_i)^{\infty}_{0}$ of length $\displaystyle
{q^r-1\over{z}}$, since
$\displaystyle{{(q^r-1)\over{z}}={{q^{r-\rho}-1\over{z}}q^{\rho}+{q^{\rho}-1\over{z}}}}$,
we put $\displaystyle{d={q^{r-\rho}-1\over{z}}}$ and
$e=\displaystyle{q^{\rho}-1\over{z}}$, then by Lemma 3, we have

$$M((s^{(\gamma,g)}_{i})^{\infty}_{0},(s^{(\gamma,g^{/})}_{i})^{\infty}_{0})\geq {4I\nu
-(I+1)Il\over{4\nu-2}}={2d\nu-\nu+2de+e\over{2\nu-1}}$$
$$=d-{\nu-2de-e-d\over{2\nu-1}}$$
$$=d-{de(z-2)\over{2\nu-1}}$$
This implies that
$$M((s^{(\gamma,g)}_{i})^{\infty}_{0},(s^{(\gamma,g^{/})}_{i})^{\infty}_{0})\geq d={q^{r-\rho}-1\over{z}}.$$

\begin{theorem}
The $\Gamma$ of (4) is a
$\displaystyle({q^r-1\over{z}},z,q^{\rho},{q^{r-\rho}-1\over{z}})$
set of frequency hopping sequence, meeting the $Peng-Fan$ bound.
\end{theorem}
\textbf{Proof:} We apply Lemma 2, where $\displaystyle I=\lfloor{\nu
z/q^{\rho}}\rfloor=q^{r-\rho}-1$,
$$(\nu-1)zH_a(\Gamma)+(z-1)z\nu H_c({\Gamma})$$
$$=({q^r-1\over{z}}-1)z{q^{r-\rho}-1\over{z}}+(z-1)z{q^r-1\over{z}}{q^{r-\rho}-1\over{z}}$$
$$=(q^r-z-1){q^{r-\rho}-1\over{z}}+(z-1)(q^r-1){q^{r-\rho}-1\over{z}}$$
$$=(q^r-2)(q^{r-\rho}-1)$$
and
$$2I\nu z-(I+1)Iq^{\rho}$$
$$=2(q^{r-\rho}-1){q^r-1\over{z}}z-q^{r-\rho}(q^{r-\rho}-1)q^{\rho}$$
$$=(q^r-2)(q^{r-\rho}-1).$$
We know that
$$(\nu-1)zHa({\Gamma})+(z-1)z\nu H_c({\Gamma})=2I\nu z-(I+1)Iq^{\rho}$$
which means that
$\displaystyle\{H_{a}({\Gamma})={q^{r-\rho}-1\over{z}},H_c({\Gamma})={q^{r-\rho}-1\over{z}}\}$
is a pair of the minimum integer solutions of the inequality
described in Lemma 2, that is, ${\Gamma}$ is a $Peng-Fan$ optimal
family of frequency hopping sequences.

\section{Conlusion}
In this paper, new optimal frequency hopping sequences are
constructed from polynomial residue class rings. When $\rho=1$, our
construction is same with the related constructions in \cite{4,5,6},
thus our construction can be take as an extension of the related
constructions in \cite{4,5,6}. Our construction posses the following
advantages:  (1) the parameters of the construction are new and
flexible, (2) by choose different parameter $\gamma$ , one can
construct many different $Peng-Fan$ optimal frequency hopping
sequence families.

\end{document}